\documentclass[12pt]{iopart}

\usepackage{cite}
\usepackage{graphicx}
\usepackage{dcolumn}

\begin{document}

\title{Eigenvalues from power--series expansions: an alternative approach}

\author{Paolo Amore\dag \ and Francisco M Fern\'andez
\footnote[2]{Corresponding author}}

\address{\dag\ Facultad de Ciencias, Universidad de Colima, Bernal D\'iaz del
Castillo 340, Colima, Colima, Mexico}\ead{paolo.amore@gmail.com}

\address{\ddag\ INIFTA (UNLP, CCT La Plata-CONICET), Divisi\'{o}n
Qu\'{i}mica Te\'{o}rica,
Diag. 113 y 64 (S/N), Sucursal 4, Casilla de Correo 16,
1900 La Plata, Argentina}\ead{fernande@quimica.unlp.edu.ar}

\begin{abstract}
An appropriate rational approximation to the eigenfunction of the
Schr\"{o}dinger equation for anharmonic oscillators enables one to obtain
the eigenvalue accurately as the limit of a sequence of roots of Hankel
determinants. The convergence rate of this approach is greater than that for
a well--established method based on a power--series expansions weighted by a
Gaussian factor with an adjustable parameter
(the so--called Hill--determinant method).
\end{abstract}

\maketitle

\section{Introduction\label{sec:intro}}

Power--series methods have proved to yield remarkably accurate eigenvalues
of simple one--dimensional and central--field quantum--mechanical models\cite
{B78,BBCK78,G82,K83,EFC87,EFC88,EFC88b,EFC90,EFC90b,AL05,K07}. There are
basically two different approaches: on the one hand, the use of Dirichlet
boundary conditions at the endpoints of a sufficiently wide interval\cite
{K83,AL05}, on the other, the Hill--determinant method and its variants\cite
{B78,BBCK78,G82,K83,EFC87,EFC88,EFC88b,EFC90,EFC90b,K07}. In this paper we
focus our attention on the latter that has been applied to a wide variety of
problems, including the vibration--rotation spectra of diatomic molecules%
\cite{EFC87,EFC88,EFC90}. The success of this method commonly depends on the
weight function which in many cases is an exponential function with a width
parameter that affects the rate of convergence of the approach\cite
{BBCK78,B78,G82}.

The purpose of this paper is to discuss an alternative approach that is
less dependent on the width parameter or scaling factor. In sec.~\ref
{sec:powerseries} we outline a well--known weighted power--series method. In
sec.~\ref{sec:Hankel} we develop an alternative approach based on a rational
approximation to that power series approach. In sec.~\ref{sec:ANHO} we apply
both approaches to the pure quartic anharmonic oscillator. In sec.~\ref
{sec:rational} we consider a rational potential with a singular point on the
complex coordinate plane. Finally, in sec.~\ref
{sec:conclusions} we draw conclusions.

\section{Weighted power--series method\label{sec:powerseries}}

Consider the Schr\"{o}dinger equation
\begin{equation}
\psi ^{\prime \prime }(x)+[E-V(x)]\psi (x)=0  \label{eq:Schro}
\end{equation}
where the potential--energy function $V(x)$ can be expanded as
\begin{equation}
V(x)=\sum_{j=1}^{\infty }v_{j}x^{2j}  \label{eq:V_series}
\end{equation}
A well known approach for the calculation of eigenvalues and eigenfunctions
is based on the ansatz
\begin{equation}
\psi (x)=e^{-ax^{2}}\sum_{j=0}^{\infty }c_{j}x^{2j+s}  \label{eq:psi_series}
\end{equation}
where $s=0$ or $s=1$ for even or odd states, respectively. If this expansion
satisfies the Schr\"{o}dinger equation (\ref{eq:Schro}), then the
coefficients $c_{j}$ are polynomial functions of the energy $E$. It has been
shown\cite{BBCK78,B78,G82} that one can obtain the allowed energies (those
consistent with square--integrable solutions) from the roots of
\begin{equation}
c_{M}(E)=0,\;M=M_{0},\,M_{0}+1,\ldots  \label{eq:cM=0}
\end{equation}
The rate of convergence of the sequence of roots $E^{[M]}$ of this equation
depends on the adjustable parameter $a$. If its value is far from optimal,
the sequences may not converge at all.

\section{The Hankel--Pad\'{e} method\label{sec:Hankel}}

The Riccati Pad\'{e} method is based on a rational approximation to the
logarithmic derivative of the eigenfunction $f(x)=s/x-\psi ^{\prime
}(x)/\psi (x)$\cite{FMT89a,FMT89b}. An appropriate truncation condition
determines the allowed energies to be the roots of Hankel determinants\cite
{FMT89a,FMT89b}. The convergence rate of the sequences of such roots is
remarkable and provides accurate eigenvalues with Hankel determinants of
relatively small dimension.

Here we explore an alternative approach based on a rational approximation to
$x^{-s}\psi (x)e^{ax^{2}}$:
\begin{equation}
\frac{\sum_{j=0}^{N+d}a_{j}x^{2j}}{\sum_{j=0}^{N}b_{j}x^{2j}}%
=\sum_{j=0}^{2N+d+1}c_{j}x^{2j}  \label{eq:Pade}
\end{equation}
Notice that we require that the Pad\'{e} approximant with just $2N+d+1$
adjustable parameters yields $2N+d+2$ coefficients of the power series (\ref
{eq:psi_series}). As in the case of the Riccati--Pad\'{e} method\cite
{FMT89a,FMT89b} it leads to a quantization condition for the energy given by
the roots of the Hankel determinants
\begin{equation}
H_{D}^{d}(E)=\left| c_{i+j+d-1}(E)\right| _{i,j=1}^{D}=0  \label{eq:Hankel_c}
\end{equation}
where $D=N+1=2,3,\ldots $, and $d=0,1,\ldots $

\section{Anharmonic oscillator\label{sec:ANHO}}

As an example consider the anharmonic oscillator
\begin{equation}
V(x)=x^{4}  \label{eq:V_x^4}
\end{equation}
The first coefficients are:
\begin{eqnarray}
c_{1} &=&a-\frac{E}{2}  \nonumber \\
c_{2} &=&\frac{a^{2}}{2}-\frac{Ea}{2}+\frac{E^{2}}{24}  \label{eq:cj_x4}
\end{eqnarray}
and the first Hankel determinant is
\begin{equation}
H_{2}^{0}=\frac{a}{30}-\frac{a^{4}}{12}-\frac{E}{60}+\frac{Ea^{3}}{6}-\frac{%
E^{2}a^{2}}{8}+\frac{7E^{3}a}{360}-\frac{E^{4}}{960}  \label{eq:H(2,0)_x4}
\end{equation}

\begin{figure}[H]
\begin{center}
\includegraphics[width=9cm]{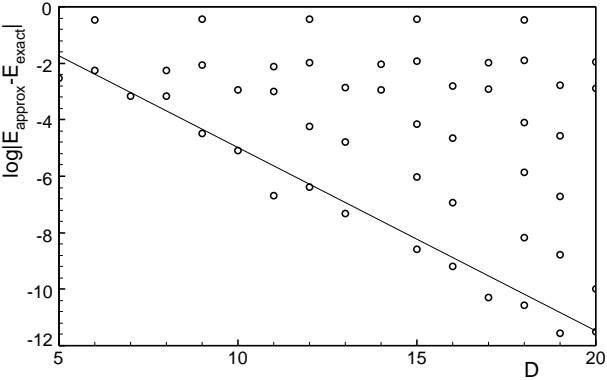}
\end{center}
\caption{Sequences of roots of $H_D^d(E)=0$ for the lowest eigenvalue
of the quartic anharmonic oscillator~(\ref{eq:V_x^4}) when $a=0$}
\label{Fig:seqs_a0}
\end{figure}

Fig.~\ref{Fig:seqs_a0} shows several sequences of roots of $H_{D}^{0}(E)=0$
when $a=0$. We appreciate that the phenomenon of multiple converging
sequences of roots present in the RPM\cite{FMT89a,FMT89b} also appears in
this case. It seems to be related to the Hankel determinant and to the
rational approximation (either to the logarithmic derivative or to just a
factor of the wave function). A straight line in Fig.~\ref{Fig:seqs_a0}
marks the sequence with the best convergence rate. One obtains similar
sequences for other values of $a$. Notice that the standard method (\ref
{eq:cM=0})\cite{B78,BBCK78,G82} does not apply to the case $a=0$.

\begin{figure}[H]
\begin{center}
\includegraphics[width=9cm]{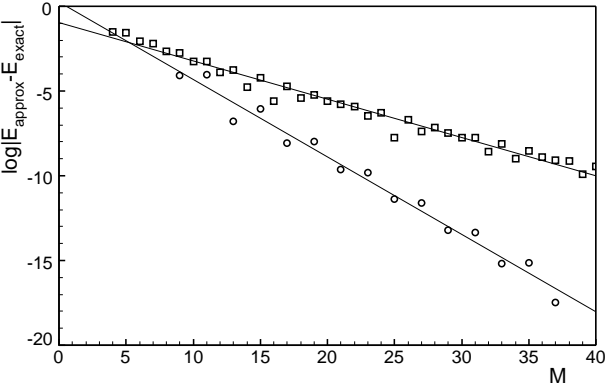}
\end{center}
\caption{Sequences of roots of $H^d_ D(E)=0$ (circles) and $c_M(E)=0$
(squares), $M=2D-1$, for the lowest eigenvalue of the quartic
anharmonic oscillator~(\ref{eq:V_x^4}) when $a=1$}
\label{Fig:seqs_a1}
\end{figure}

Fig.~\ref{Fig:seqs_a1} shows the logarithmic error $\log
|E_{approx}-E_{exact}|$ of the sequences of roots of $H_{D}^{0}(E)=0$ and $%
c_{M}(E)=0$ for $a=1$ in terms of the number of coefficients $M=2D-1$
required by the calculation. Straight lines show the overall trend of the
logarithmic sequences. We appreciate that the Hankel sequence converges
faster than the one for the standard approach. The ``exact'' result
$E_{exact}=1.0603620904841828996$ is simply a more accurate estimate
of the eigenvalue provided
by the RPM\cite{FMT89a,FMT89b}.

\begin{figure}[H]
\begin{center}
\includegraphics[width=9cm]{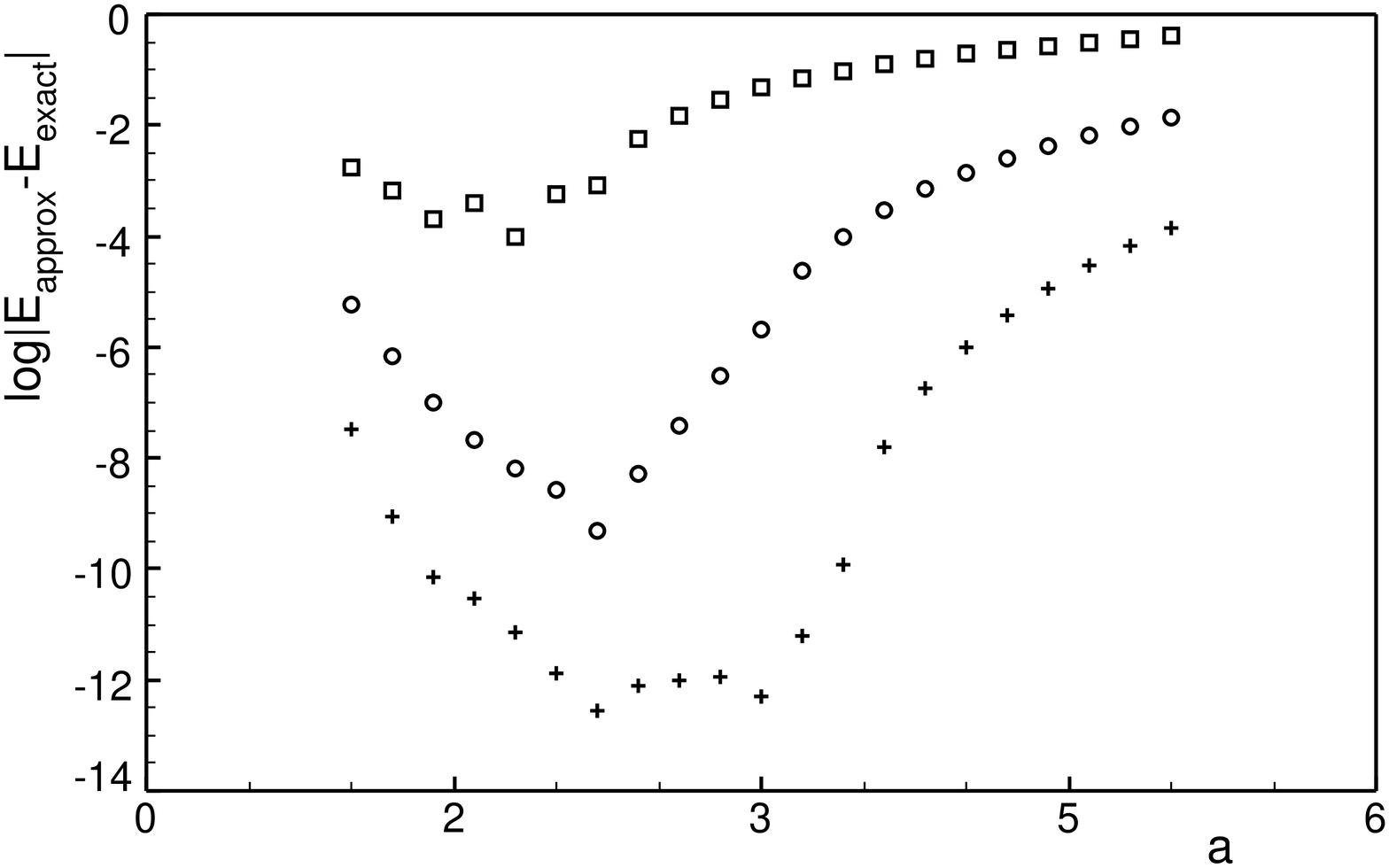}
\end{center}
\caption{Logarithmic error of the roots of $c_M(E)=0$ as function of $a$ for
$M=9$ (squares), $M=19$ (circles) and $M=29$ (crosses) for the lowest
eigenvalue of the quartic anharmonic oscillator~(\ref{eq:V_x^4})}
\label{Fig:var_a}
\end{figure}

Fig~\ref{Fig:var_a} shows the variation of the logarithmic error of the
roots of $c_{M}(E)=0$ with $a$ for three values of $M$. We appreciate that
the optimal value of the adjustable parameter for the quartic oscillator (%
\ref{eq:V_x^4}) is about $a\approx 2.5$.

It is not necessary to have the ``exact'' energy in order to estimate an
optimal value of the adjustable parameter. If $E_{n}^{[k]}$ is the
approximation of order $k$ to the $n$--th eigenvalue, we simply monitor the
convergence of the sequence in terms of, for example, $\log
|E_{n}^{[k+1]}-E_{n}^{[k]}|$, where $k=k_0,k_0+1,\ldots$.

\section{Rational potential\label{sec:rational}}

Both the straightforward power--series method and the Hankel--Pad\'{e}
approach apply successfully to polynomial potentials as illustrated in the
preceding section by means of a simple nontrivial example. In what follows
we consider the rational potential
\begin{equation}
V(x)=x^{2}+\lambda x^{2}/(1+gx^{2}),\,-\infty< \lambda < \infty ,\,g>0  \label{eq:V_rational}
\end{equation}
that has been studied by several authors\cite{M78,K79,BB80,F81,H81,V81,F82,H82,LL82,
WWN82,BBH83,CM83,H83,Z83a,Z83b,C84,Z84,FV85,H85,MP85,FDV86,BL87,FV87,RR87,V87,G88,
H88,RRV88,RRR88,SRL88,VD89,
BV89,L89,H90,RR90,ADV91,F91,PM91,W91,AV93,HHSJS93,SG95,I02,SHC06}. Among the
approaches applied to this model we mention perturbation
theory\cite{BBH83,K79,LL82,W91,Z84}, including the $1/N$
expansion\cite{RRR88,SRL88,V87}, variational
methods\cite{M78,BB80,FDV86,SG95,SHC06}, and in particular the Rayleigh--Ritz
method\cite{M78,FDV86,SG95,SHC06}. One can easily obtain exact solutions
to the Schr\"odinger
equation with the potential~(\ref{eq:V_rational}) for some values of the parameters
$\lambda$ and $g$\cite{ADV91,VD89,BL87,BV89,CM83,F81,F82,G88,H82,H83,H90,LL82,
L89,PM91,RR87,RRV88,RR90,SHC06,SRL88,V81,WWN82,Z83a,Z83b} that prove suitable
for testing approximate methods.

The power series (\ref{eq:V_series}) converges only for
$|x|<|x_{g}|$, where $x_{g}=\pm i/\sqrt{g}$ are the two poles of the
potential--energy function on the imaginary axis of the complex $x$--plane.
If $g\ll 1$ the
eigenfunction is negligible for $|x|>|x_{g}|$ and the Hill--determinant
method may yield reasonable results for the lowest energies and only after
judicious truncation of the sequences of roots\cite{EFC87}.
On the other hand, the expansion of the
wavefunction in a power
series of the variable $u=x^{2}/(1+gx^{2})$ leads to a successful approach
for all values of $g$ and $\lambda $\cite{F91}.

Since the Hankel--Pad\'{e} method is based on a rational approximation to
the wavefunction, one expects that it takes into account the singularities
properly, succeeding even for moderate values of $g$.
In what follows we compare it with the weighted power--series method (\ref
{eq:psi_series}). First of all, notice that $V(x)/x^{2}\rightarrow 1$ as
$|x|\rightarrow \infty $ so that we expect $a=1/2$ to be optimal\cite{G82}
and choose this width--parameter value from now on. We have verified that the
Hankel--Pad\'e approach yields reasonable results for other values of $a$ such as,
for example, $a=0$, $a=1$ and $a=3/2$.

Table~\ref{tab:rational1} shows the results of the Hankel--Pad\'{e}
calculation of the ground--state eigenvalue of the Schr\"{o}dinger
equation (\ref{eq:Schro}) with the rational potential (\ref{eq:V_rational})
for $\lambda =1$ and three values of $g$. Notice that present Hankel--Pad\'{e}
results are more accurate than those obtained earlier by means of the
Rayleigh--Ritz variational method\cite{M78}, and comparable to those
provided by a kind of iterative solution of the Rayleigh--Ritz secular equation
with an adjustable parameter\cite{FDV86}. There are much more accurate results
in the literature; for example, Stubbins and Gornstein\cite{SG95} obtained
$E_0=1.38053180093804523438995006009$ and $E_0=1.23235072340605781386206995868$
for $g=0.1$ and $g=1$, respectively.

Table~\ref{tab:rational2} shows results from the
Hill--determinant method. A lack of entry means that we did not find any
root in the interval $0.5<E<1.5$. Notice that while the Hankel--Pad\'{e}
approach converges smoothly the Hill--determinant method does not, even for
$g=0.1$. Besides, the latter approach does not give any reasonable result for
$g=1$. For $g=0.1$ the roots of the Hill--determinant oscillate about the
exact eigenvalue, giving the tightest bounds for $9\leq M \leq 12$, before the sequence
begins to diverge. Averaging the roots for $M=10$ and $M=11$ one estimates
$E_0=1.38053181$ that is quite close to the exact eigenvalue. However, this
strategy is only practical for sufficiently small values of $g$ as discussed above.
We have thus verified our earlier supposition that the Hankel--Pad\'{e} method
should correct a possible failure of the power--series approach caused by
singular points of the potential--energy function in the complex coordinate
plane.

\section{Further comments and conclusions\label{sec:conclusions}}

Clearly, the results of the preceding sections show that

\begin{itemize}
\item  The sequence of roots of the Hankel determinants (\ref{eq:Hankel_c})
converges more smoothly than the sequence of roots of equation (\ref{eq:cM=0}%
) for polynomial potentials.

\item  The rate of convergence of the sequence of roots of the Hankel
determinants (\ref{eq:Hankel_c}) is not so strongly dependent on the value
of $a$ as the sequence of roots of the standard approach (\ref{eq:cM=0}). In
fact, the former converges where the latter does not (even when $a=0$).

\item  The Hankel--Pad\'{e} method is preferable for the treatment of
potential--energy functions with singularities in the complex coordinate plane
that limit seriously the range of applicability of the power series.

\item  However, from a purely practical point of view it is worth noticing
that when both approaches are successful, the calculation of the roots of
the Hankel determinants typically requires more CPU time.

\item  There is more than one sequence of roots of the Hankel determinant (%
\ref{eq:Hankel_c}) that converges towards a given eigenvalue. Present
approach shares this curious phenomenon with the RPM\cite{FMT89a,FMT89b} and
appears to be a feature of the Hankel determinants constructed from the
coefficients of the power series coming from either the Riccati equation or
the Schr\"{o}dinger one.
\end{itemize}

The Schr\"{o}dinger equation with the simple potential--energy functions
discussed above can easily be treated by means of the Rayleigh--Ritz
variational method and the basis set of eigenfunctions $\{\phi _{n}\}$ of
the harmonic oscillator $\hat{H}=\hat{p}^{2}+\omega ^{2}x^{2}$, where $%
\omega $ is an adjustable parameter. The problem reduces to the
diagonalization of the Hamiltonian matrix $\mathbf{H}$ with elements $%
H_{ij}=\left\langle \phi _{i}\right| \hat{H}\left| \phi _{j}\right\rangle $.
The main advantage of this approach is that it provides upper bounds to all
the eigenvalues\cite{M78,FDV86,SG95,SHC06}. Besides, in some cases $H_{ij}=0$
for all $|i-j|>k$, and the resulting secular equation with a band matrix can be
treated as a recurrence relation. In this way one does not have to
diagonalize a large matrix but simply to find the roots of a determinant of
much smaller constant dimension\cite{FOT86}. This is precisely the case for the
simple examples discussed above. However, this variational method may not be
practical if the calculation of the matrix elements of the potential--energy
function is too difficult. In that case the power--series methods and its
variants may be preferable.

\begin{table}[tbp]
\caption{Hankel--Pad\'e estimate of the ground--state eigenvalue of the
Schr\"{o}dinger equation with the rational potential (\ref{eq:V_rational})
for $\lambda=1$}
\label{tab:rational1}
\begin{center}
\par
\begin{tabular}{D{.}{.}{8}D{.}{.}{20}D{.}{.}{20}D{.}{.}{20}}
\hline
\multicolumn{1}{c}{$D$}& \multicolumn{1}{c}{$g=0.1$} & \multicolumn{1}{c}{$g=0.2$}
 & \multicolumn{1}{c}{$g=1$}  \\

\hline
2  & 1.385                  & 1.353120                &   1.21                      \\
3  & 1.380525               & 1.353123                &   1.23                      \\
4  & 1.3805318              & 1.3529481               &   1.232                      \\
5  & 1.3805322              & 1.3529489               &   1.2323                      \\
6  & 1.38053181             & 1.352948023             &   1.23234             \\
7  & 1.3805318009377        & 1.352952                &   1.232348             \\
8  & 1.380531800938043      & 1.352948022755          &   1.2323502             \\
9  & 1.3805318009380452     & 1.352948037359          &   1.2323506             \\
10 & 1.380531800938045232   & 1.352948022753577       &   1.23235069            \\
11 & 1.3805318009380452345  & 1.352948022753566       &   1.23235072            \\
12 & 1.3805318009380452344  & 1.35294802275357088     &   1.232350721            \\
13 & 1.3805318009380452344  & 1.35294802275357081     &   1.232350723            \\
14 &                        & 1.3529480227535708289   &   1.2323507233            \\
15 &                        & 1.3529480227535708284   &   1.23235072337              \\
16 &                        & 1.3529480227535708285   &   1.23235072339              \\
17 &                        & 1.3529480227535708284   &   1.232350723403              \\
18 &                        & 1.3529480227535708284   &   1.232350723405              \\
19 &                        &                         &   1.2323507234057            \\
20 &                        &                         &   1.23235072340595        \\
21 &                        &                         &   1.23235072340602        \\
22&&                                                  &   1.232350723406047  \\
23&&                                                  &   1.232350723406054  \\
24&&                                                  &   1.2323507234060566 \\
25&&                                                  &   1.2323507234060574
\end{tabular}
\par
\end{center}
\end{table}

\begin{table}[tbp]
\caption{Hill--determinant estimate of the ground--state eigenvalue of the
Schr\"{o}dinger equation with the rational potential (\ref{eq:V_rational})
for $\lambda=1$}
\label{tab:rational2}
\begin{center}
\par
\begin{tabular}{D{.}{.}{1}D{.}{.}{6}D{.}{.}{6}}
\hline
\multicolumn{1}{c}{$M$}& \multicolumn{1}{c}{$g=0.1$} & \multicolumn{1}{c}{$g=0.2$}
   \\

\hline
2  & 1.59   & 1.59               \\
3  & 1.32   & 1.26               \\
4  & 1.41   & 1.43                \\
5  & 1.37   & 1.30                 \\
6  & 1.389   & 1.42         \\
7  & 1.375   & 1.29          \\
8  & 1.385   & 1.46           \\
9  & 1.377   & 1.22           \\
10 &  1.384  & 1.82          \\
11 &  1.377  & 1.03          \\
12 &  1.384  &                          \\
13 &  1.376  &                          \\
14 &  1.386  &                           \\
15 &  1.373  &                              \\
16 &  1.391  &                              \\
17 &  1.364  &                               \\
18 &  1.409  &                               \\
19 &  1.337  &                              \\
20 &  1.48  &                              \\
21 &  1.25

\end{tabular}
\par
\end{center}
\end{table}

\end{document}